\documentclass[aps,pre,groupedaddress,twocolumn]{revtex4}
 \pdfoutput=1

\usepackage{ifpdf}

\ifpdf
\usepackage[pdftex]{graphicx}
\else
\usepackage{graphicx}
\usepackage{bm}
\fi
\usepackage{hyperref}

\usepackage{subfigure}

\begin{document}

\ifpdf
\DeclareGraphicsExtensions{.pdf, .jpg}
\else
\DeclareGraphicsExtensions{.eps, .jpg}
\fi

\def\hslash{\hbar}
\def\imag{i}
\def\grad{\vec{\nabla}}
\def\div{\vec{\nabla}\cdot}
\def\curl{\vec{\nabla}\times}
\def\DDt{\frac{d}{dt}}
\def\ddt{\frac{\partial}{\partial t}}
\def\ddx{\frac{\partial}{\partial x}}
\def\ddy{\frac{\partial}{\partial y}}
\def\lap{\nabla^{2}}
\def\divv{\vec{\nabla}\cdot\vec{v}}
\def\gradS{\vec{\nabla}S}
\def\vvec{\vec{v}}
\def\wc{\omega_s}
\def\<{\langle}
\def\>{\rangle}
\def\Tr{{\rm Tr}}
\def\Csch{{\rm csch}}
\def\Coth{{\rm coth}}
\def\Tanh{{\rm tanh}}
\def\g2{g^{(2)}}
\newcommand{\al}{\alpha}

\newcommand{\la}{\lambda}
\newcommand{\del}{\delta}
\newcommand{\om}{\omega}
\newcommand{\ep}{\epsilon}
\newcommand{\pd}{\partial}
\newcommand{\bra}{\langle}
\newcommand{\ket}{\rangle}
\newcommand{\bbra}{\langle \langle}
\newcommand{\kket}{\rangle \rangle}
\newcommand{\non}{\nonumber}

\newcommand{\be}{\begin{eqnarray}}
\newcommand{\ee}{\end{eqnarray}}

\newcommand{\bea}{\begin{eqnarray}}
\newcommand{\eea}{\end{eqnarray}}

\title{Electron transfer dynamics using projected modes}



\author{Andrey Pereverzev}
\email[email:]{Andrey.Pereverzev@mail.uh.edu}
\affiliation{Department of Chemistry and Texas Center for Superconductivity, 
University of Houston \\ Houston, TX 77204}

\author{Eric R. Bittner}
\email[email:]{bittner@uh.edu}
\thanks{John Simon Guggenheim Fellow, 2007-8}
\affiliation{Department of Chemistry and Texas Center for Superconductivity, 
University of Houston \\ Houston, TX 77204}

\date{\today}

\begin{abstract}
For electron-phonon Hamiltonians with the  couplings linear in the phonon operators we construct a class of unitary transformations that  separate the normal modes into two groups. The modes in the first group  interact with the electronic degrees of freedom directly. The modes in the second group interact directly  only with the modes in the first group but not with the electronic system. We show that for the $n$-level electronic system the minimum number of modes in the first group is $n_s=(n^2+n-2)/2$. The separation of the normal modes into two groups allows one to develop new approximation schemes. We apply one of such schemes to study exitonic relaxation in a model semiconducting molecular heterojuction.
 \end{abstract}


\maketitle
\section{Introduction}

One of the central assumptions in formulating the relaxation dynamics
of a quantum system embedded in a condensed media is the partitioning
between degrees of freedom treated explicitly and implicitly.  Often,
we are guided by our ``physical intuition'' in separating the system
(explicit degrees of freedom) from the bath (implicit degrees of
freedom).  This intuitive approach breaks-down severely when the
coupling between system and bath degrees of freedom become too strong.
This situation is exasperated when one attempts to treat some of the
degrees of freedom using quantum mechanics and the remaining degrees
of freedom using classical mechanics.  What is clearly lacking is a
unambiguous approach for formally partitioning a system with both
discrete and continuous degrees of freedom into interacting sub-spaces
such that the motions that are most strongly coupled are given
explicit treatment while the remainder serve as auxiliary degrees of
freedom (quantum or classical), as a thermal bath, or are ignored
completely.  Clearly, such a partitioning is dependent upon the
specific system at hand.  However, the prescription for making the
partitioning should be universal and depend only upon the form of the
original Hamiltonian.

In many respects, we anticipate the presence of ``special'' degrees of
freedom.  One of the central tenants of modern chemical physics is
that of the reaction coordinate, espectially in the Marcus-Hush model
of charge-transfer between donor and acceptor species.  The reaction
coordinate that parameterizes the free-energy parabolas in this
classic model are not explicitly defined in terms of actual nuclear
motions.  The driving force, electronic coupling, and reorganization
energy terms appearing in the Marcus-Hush expression for the
electron-transfer rate are all experimentally derived quantities.
Nuclear motion enters into the description when we make the connection
between the polarization field and the reorganization and relaxation
of the donor and acceptor molecules and their surroundings.

In this paper we present a procedure for deducing the reaction
coordinate for a multidimensional electron transfer process.  By using
a series of canonical transformations
 we can uniquely
decompose the fully-coupled problem into a subset of strongly
correlated vibronic motions that have clear dynamical and structural
meaning and carry all of the electron/phonon couplings.  This subspace
is coupled to the remaining vibrational degrees of freedom which in
turn are completely decoupled from the electron transfer dynamics, but
serve as a dissipative subsystem. 

\section{Projected modes}

\begin{widetext}

A wide class of the electron-phonon systems is described by the following Hamiltonian 
\be
H=\sum_{a=1}^{n}\epsilon_a |a\ket\bra a|+
\sum_{a=b=1}^{n}\sum_{i=1}^{N}g_{abi}x_i|a\ket\bra b|
+\sum_{i=1}^{N}\frac{p_i^2}{2}+\sum_{i=1}^{N}\frac{\om_i^2x_i^2}{2}. \label{Ham}
\ee
Here $|a\ket$'s denote electronic states with energies $\epsilon_a$,
$x_i$ and $p_i$ are coordinate and momentum operators for the normal
mode $i$ with frequency $\omega_i$, and $g_{abi}$ are the coupling parameters
of the electron-phonon interaction. $N$ is the number of normal modes and 
$n$ is the number of electronic states. We seek to rewrite $H$ in terms of the new
coordinate $X_i$ and momentum $P_i$ operators  so that it has the following
form
\be
H=\sum_{a=1}^{n}\epsilon_a |a\ket\bra a|+
\sum_{a=b=1}^{n}\sum_{i=1}^{n_s}g'_{abi}X_i |a\ket\bra b|
+\sum_{i=1}^{n_s}\sum_{j=n_s+1}^N \gamma'_{ij}X_iX_j+\sum_{i=1}^{N}\frac{{P}_i^2}{2}+
\sum_{i=1}^{N}\frac{{\omega'}_i^2{X}_i^2}{2}. \label{Ham-prime}
\ee
where
$n_s=\frac{n(n+1)}{2}$ and  $\omega'_i$ are the new oscillator 
frequencies. In the new Hamiltonian only $n_s$ modes are directly coupled to
the electronic degrees of freedom with $g'_{abi}$ being the new coupling parameters. 
We will refer to these modes as the system modes \cite{gindensperger:144103, gindensperger:144104}.
These $n_s$ modes are also coupled to the
$N-n_s$ remaining normal modes (that are not directly coupled to the electronic 
degrees of freedom). We will call these $N-n_s$ modes the bath modes. Coefficients
$\gamma'_{ij}$ give the couplings
between the system modes and the bath modes.
The transition form Eq. (\ref{Ham}) to Eq. (\ref{Ham-prime}) can be achieved
by a suitable linear transformation of the coordinate and momentum operators.

Consider the following linear transformations
\bea
X_i&=&\sum_jM_{ij}x_j, \label{transx} \\ 
P_i&=&\sum_j (\widetilde{M}^{-1})_{ij}p_j. \label{transp}
\eea
Here $M_{ij}$ are the elements of an arbitrary real matrix ${\bf M}$.  We will assume that ${\bf M}$ has
an inverse, ${\bf M}^{-1}$. ${\bf \widetilde{M}^{-1}}$ denotes the transpose of ${\bf M}^{-1}$.
It can be shown that transformations (\ref{transx}) and (\ref{transp})  
are canonical, i. e., $X_i$  and  $P_i$  satisfy the usual commutation relations 
for coordinates and momenta. Since matrix ${\bf M}$ is real the new operators $X_i$  and  $P_i$ are Hermitian.
The inverse transformations are given by 
\bea
x_i&=&\sum_jM^{-1}_{ij}X_j, \label{transinverx}\\ 
p_i&=&\sum_j \widetilde{{M}}_{ij}P_j. \label{transinverp}
\eea
In Appendix A we show that transformations (\ref{transx}) and (\ref{transp}) 
can always be viewed as unitary transformations of
operators $x_i$  and  $p_i$. 

Our goal of going to the new coordinates can be achieved with orthogonal matrices ${\bf M}$. 
Therefore, in this section we
will assume that  ${\bf M}$ is orthogonal, i. e., $\widetilde{\bf M}={\bf M}^{-1}$.
We will discuss the possible uses of the non-orthogonal matrices below.

In order to obtain the explicit form of ${\bf M}$ we substitute expressions (\ref{transinverx}) and (\ref{transinverp}) for $x_i$ and $p_i$ into
Eq. (\ref{Ham}). This gives
\be
H=\sum_{a=1}^{n}\epsilon_a |a\ket\bra a|+
\sum_{a=b=1}^{n}\sum_{i=1}^N\sum_{j=1}^{N}g_{abi}M^{-1}_{ij}X_j|a\ket\bra b|
+\sum_{i=1}^{N}\frac{P_i^2}{2}+\sum_{i=j=1}^{N}\frac{w_{ij}X_iX_j}{2}. \label{Ham2}
\ee
Here 
\be
w_{ij}=\sum_k M_{ik}\om^2_kM^{-1}_{kj}.
\ee
Note that the momentum operators remain uncoupled for any orthogonal ${\bf M}$ 
in Eq. (\ref{transinverp})
because the matrix associated with the quadratic form $\sum_i p_i^2$ is the unit matrix and is invariant 
under the orthogonal transformations.
\end{widetext}

Hamiltonian (\ref{Ham2}) will have the form (\ref{Ham-prime}) if the following
requirements are satisfied. First,
the summation over $j$ in the second term has to extend only from $1$ to $n_s$. 
Second, the mode-mode
interaction part for the coordinate operators must be such that there is no interaction within the group of 
the $n_s$ system modes and no interaction within the group of the $N-n_s$ bath modes while 
the interaction between these to groups remains.

Consider the second term of Eq. (\ref{Ham2}). It contains the sum  
$\sum_{ij}g_{abi}M^{-1}_{ij}X_j$. This expression involves 
multiplications from the left of the row vector ${\bf g}_{ab}$ with elements $g_{abi}$ on 
the $N\times N$  matrix ${\bf M}^{-1}$. Since $g_{abi}=g_{bai}$ there are $n_s=n(n-1)/2$ linearly
independent vectors ${\bf g}_{ab}$ for the $n$ level electronic system. To simplify the notation 
we will denote these
$n_s$ independent vectors ${\bf g}_{\alpha}$ where $\alpha$ can run from $1$ to $n_s$.
Recall that the columns of any orthogonal matrix form
a complete set of orthonormal vectors. 
Then, our first requirement of the reduction of the number of operators $X_i$ appearing 
in the electron-phonon interaction to
$n_s$ can be viewed as the search for orthogonal matrices ${\bf M}^{-1}$ 
whose $N-n_s$ columns  
are orthogonal to $n_s$ vectors ${\bf g}_{\alpha}$. Obviously these matrices are not unique. 
To obtain the columns of such 
matrices we construct
 $n_s$ arbitrary orthonormal vectors in the subspace of  
$n_s$ vectors ${\bf g}_{\alpha}$ and $N- n_s$ arbitrary orthonormal vectors
in the subspace that is orthogonal to all ${\bf g}_{\alpha}$'s. We will call these two groups
${\bf k}^s_{\alpha}$ and ${\bf k}^b_{\alpha}$, respectively. 

The unique choice of matrix ${\bf M}^{-1}$ 
is achieved by satisfying our second requirement. Explicitly, vectors ${\bf k}^s_{\alpha}$ and ${\bf k}^b_{\alpha}$
must be such that the $n_s\times n_s$ Hessian matrix for the system oscillators
\be
W^s_{\alpha \alpha'}=\sum_{i}^N k^s_{\alpha i}\omega_i^2k^s_{\alpha'i}
\ee
and $(N-n_s)\times(N-n_s)$ Hessian matrix for the bath oscillators
\be
W^b_{\alpha \alpha'}=\sum_{i}^N k^b_{\alpha i}\omega_i^2k^b_{\alpha'i}
\ee
 are both diagonal. The eigenvalues of these matrices will give the squares of the new oscillator frequencies.
The coupling coefficients $\gamma'_{\alpha \alpha'}$ in Eq. (\ref{Ham-prime}) are given by 
\be
\gamma'_{\alpha \alpha'}=\sum_{i}^N k^s_{\alpha i}\omega_i^2k^b_{\alpha'i}.
\ee
The first subscript in $\gamma_{\alpha \alpha'}$ refers to the system modes and the second one to
the bath modes.
Coefficients $g'_{a b \alpha}$ for the electron-phonon coupling in Eq. (\ref{Ham-prime}) are 
\be
g'_{ab\alpha}=\sum_i g_{abi} k^s_{\alpha i}.
\ee

An alternative view on the above derivation can be obtained if we introduce a
projection matrix onto the subspace of vectors ${\bf g}_{\alpha}$. 
  We do so by defining the projection operator
\begin{eqnarray}
{\bf P} = \sum_{\alpha\beta}'S^{-1}_{\alpha\beta} {\bf g}_\alpha \otimes {\bf g}_\beta
\end{eqnarray}
with $S_{\alpha\beta} ={\bf g}_{\alpha}\cdot {\bf g}_{\beta} $
and $\otimes$ denotes the outer product.  The summation is
restricted to only linearly-independent vectors. 
This is an $N\times N$ matrix projects out all primitive modes that 
are directly coupled to the electronic degrees of freedom 
 and its complement ${\bf Q} ={\bf I}-{\bf P}$ projects out all motions 
not drectly coupled.  It is simple to show that ${\bf P}\cdot {\bf Q} = 
{\bf Q}\cdot {\bf P}$.

We can now form the matrix ${\bf M}^{-1}$ that transforms between the
primitive  and projected coordinates by finding the eigenvectors of the
block-diagonal elements of the $N\times N$ Hessian matrix for the
primitive modes following the projection
 \begin{eqnarray} {\bf K}={\bf P}{\bf \om}^2{\bf P} 
+ {\bf Q}{\bf\om}^2{\bf Q}.  
\end{eqnarray}
 Both ${\bf P}{\bf \om}^2{\bf P}$ and ${\bf Q}{\bf \om}^2{\bf Q}$ are
 $N\times N$ matrices.  However, the first will only have $n_{p}$ non-zero
 eigenvalues $\{\omega_s^{2}\}$ with corresponding eigenvectors
 $\{M_s\}$ while the second (residual set) will have $N-n_s$ non-zero eigenvalues
 $\{\omega_{r}^{2}\}$ with corresponding eigenvectors $\{M_{r}\}$.
 Thus the full $N\times N$ transformation matrix is formed by joining  the 
non-trivial vectors from the $P$ subspace to the $N-2$ vectors from the
$Q$ subspace
$M =\{M_s,M_{r}\}$.

The coupling between the ${\cal P}$ and ${\cal Q}$ subspaces,
$\gamma'_{\alpha\alpha'}$, are then given by the non-diagonal blocks
of the hessian
\begin{eqnarray} {\bf L}={\bf P}{\bf \omega}^2{\bf Q} + {\bf Q}{\bf
  \om}^2{\bf P} 
\end{eqnarray}
 transformed into the eigenbasis of $K$. 
\begin{eqnarray}
{\bf\gamma'}= ({\bf M}_s\cdot{\bf L}\cdot \widetilde{\bf M}_r), 
\end{eqnarray}
where ${\bf M}_s$ is a $n_s\times N$ matrix whose rows are vectors $\{M_s\}$ and
${\bf M}_r$ is a $N-n_s\times N$ matrix whose rows are vectors $\{M_{r}\}$ 
 Finally, the
electron/phonon coupling coefficients $g'_{abi}$ are obtained as
\begin{eqnarray}
 {\bf g}'_{ab}={\bf g}_{ab}\cdot {\bf M}_s.  
\end{eqnarray} 
In short, the projection scheme based upon the linear coupling
coefficients provides a robust and efficient way to extract a subset
of motions that are strongly coupled to the electronic degrees of
freedom. 

The final hessian matrix now has the form:
\begin{eqnarray}
\tilde{K} = \left[
\begin{array}{cc}
 \omega_s^{2}  &  \gamma            \\
\gamma              &  \omega_{r}^{2}
\end{array}
\right],
\end{eqnarray}
where $\omega_s$ are the frequencies of the $n_s$ coupling modes, 
$\omega_{r}$ are the frequencies of the residual $N-n_{r}$ modes. The two
types of modes are coupled via $\gamma$.   One can easily verify that 
this new hessian is related by orthogonal transformation to the original (diagonal)
hessian matrix and obtain two sets of Boson
 operators $A_{i}\in {\cal P}$ with $[A_s,A_{c'}^{\dagger}]=\delta_{cc'}$ and
 $B_{r}\in{\cal Q}$ with  $[B_{r},B_{r'}^{\dagger}]=\delta_{rr'}$
 which separately act in the ${\cal P}$ and ${\cal Q}$ subspaces respectively so that
 $[A_s,B_{r'}] = 0$. Clearly, modes in ${\cal Q}$ are orthogonal but coupled to the modes in ${\cal P}$. 

Several generalizations of the transformations considered above are possible.
Any Hamiltonian in which the  electron-phonon coupling involves 
$n_s$ sums of the form $\sum_i a_ix_i$ or $\sum_i b_ip_i$ where $a_i$ and $b_i$ are constants can be
 transformed in such a way that only $n_s$ 
 modes are directly coupled to the electronic degrees of freedom.
An example of such Hamiltonian can be found in Ref. \cite{pereverzev:104906}. Note that it is also possible 
to further reduce the number of the system modes by one, if the completeness relationship for the electronic degrees of freedom is used. 
So far we considered the unitary transformation that isolates the minimum number of modes coupled to the
electronic system. 
Imposing different requirements on  the unitary transformation we can obtain different  decompositions of the phonons into groups. Certain schemes may be better suited to 
a particular problem than others. 
For example, in  Ref. \cite{Cederbaum:2005,gindensperger:144103,gindensperger:144104}
 the so-called Mori chain transformation was applied  in the case of two electronic states.  Here, 
 the primitive Hamiltonian is first transformed as above so that only three normal modes are coupled to the two-level system and the bath contains $N-3$ modes.  
 However, at the next stage a new transformation is applied to the bath modes so that the three system modes are coupled directly to only three other bath modes while the new bath consists 
of $N-6$ modes. The procedure is iterated until all the modes are arranged into a hierarchy whereby each level contains
three modes coupled to the three modes in the previous level and three modes in the next level. Note that the modes within each level of the hierarchy are not coupled  direct 
to each other but only indirectly via the  neighboring levels of the chain. 
We recently applied this  type of transformation to study exciton relaxation in a conjugated polymers heterojunction Refs. \cite{tamura:021103,tamura:107402,Tamura:2006} and in the
photoreactive yellow protein\cite{GromovE.V.ja069185l}.
Alternative form of the Mori chain can be constructed. It is possible to develop a hierarchy in which modes within each level are coupled to each other but each of them is coupled to only one mode from the previous level and one mode from the next. Without going into details we note that to  construct such a chain requires using non-orthogonal matrices
 whose rows involve vectors formed from coupling coefficients without preliminary orthogonalization as above. Note, however, that the new operators are still related to the old ones through a unitary transformation as shown in Appendix A.

In Appendix B, we show that  that it is possible to transform 
the Hamiltonian in Eq. \ref{Ham-prime} to the form in which there are still $n_s$ system modes but  each of the operators $|\alpha\ket\bra \beta|+|\beta\ket\bra \alpha|$ (for $\alpha\neq \beta$) and $|\alpha\ket\bra \alpha|$ interacts with its own single  mode.  In particular, for the two-level system only one mode is directly responsible for the electronic transitions.

\section{Electronic relaxation in a molecular heterojunction dimer} \label{example}

\begin{figure}
\includegraphics[width=\columnwidth]{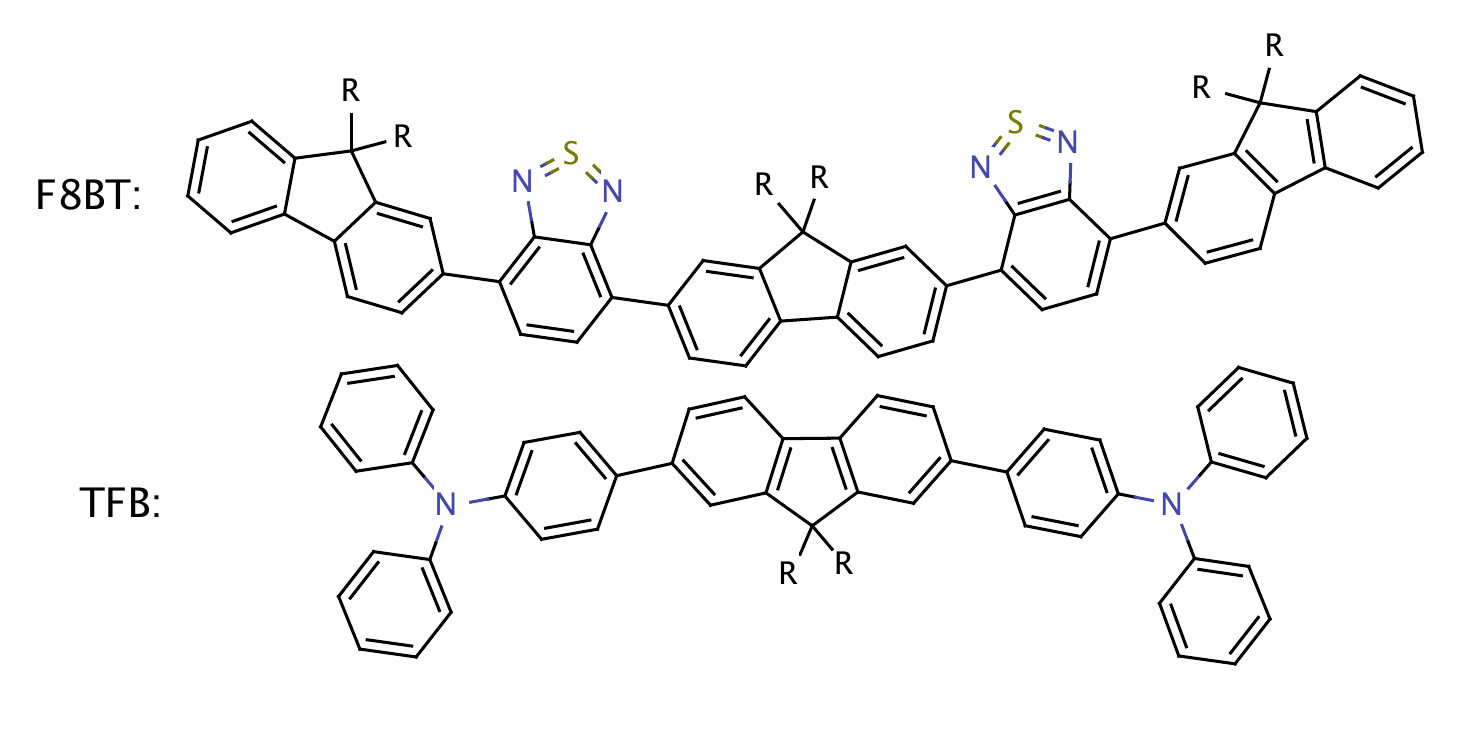}
\caption{Chemical structures of F8BT and TFB oligomers. }\label{dimers}
\end{figure}

As an application of our approach we consider the decay of an excitonic state into a charge-separated
state at the interface between two semiconducting polymer phases (TFB) and  
(F8BT). The chemical structures of these are given in Fig.~\ref{dimers}.   Such materials have been extensively studied for their potential in 
organic light-emitting diodes and organic photovoltaics
\cite{morteani:1708,morteani:163501,dhoot:2256,morteani:247402,morteani:244906,sreearunothai:117403}. 
At the phase boundary, the material forms a type-II semiconductor
heterojunction with the off-set between the valence bands of the two
materials being only slightly more  than the binding energy of 
an exciton placed on either the TFB or F8BT polymer.  As a result, an exciton on 
the F8BT side will dissociate to form a charge-separated (exciplex) state at the 
interface. \cite{morteani:244906,morteani:1708,morteani:247402,silva:125211,stevens:165213}
$$
{\rm F8BT}^*:{\rm TFB}\longrightarrow {\rm F8BT}^-:{\rm TFB}^+.
$$
Ordinarily, such type II systems are best suited for photovoltaic rather than LED 
applications
However, LEDs fabricated from phase-segregated 50:50 blends of TFB:F8BT 
give remarkably efficient electroluminescence efficiency due to {\em secondary} 
exciton formation due the back-reaction
$$
{\rm F8BT}^-:{\rm TFB}^+\longrightarrow {\rm F8BT}^*:{\rm TFB}.
$$ 
as thermal equilbrium between the excitonic and charge-transfer
states is established.  This is evidenced by long-time emission,
red-shifted relative to the emission from the exciplex, accounting
for nearly 90\% of the integrated photo-emission.

Here we consider only two  electronic levels corresponding to 
$$
|XT\rangle =   {\rm F8BT}^*:{\rm TFB} \,\,\&\,\, |CT\rangle= {\rm F8BT}^-:{\rm TFB}^+.
$$ 
We take the vertical energies for these two states as $\epsilon_{CT} = 2.019$eV and
$\epsilon_{XT}= 2.306$ eV. 
As was shown in Ref. \cite{khan:085201,bittner:214719,karabunarliev:5863,karabunarliev:11372}, 
in  poly-fluorene based systems, 
there are essentially two phonon bands that are coupled strongly
to the electronic degrees of freedom as evidenced by their presence as
vibronic features in the vibronic emission spectra.  Within our model these are 
represented by two narrow sine-shaped bands.  The 
first, centered about 97.40 cm$^{-1}$ with a width of 
10.16 cm$^{-1}$, 
accounts for the low-frequency torsional motions while the second,
higher frequency band centered at  1561.64 cm$^{-1}$ with width 156.30 cm$^{-1}$
 accounts for the  C=C ring-stretching modes.   These primitive modes are entirely localized on 
 one chain or the other. 
The vibronic couplings within the model were determined by 
comparison between the Franck-Condon peaks of the predicted and observed spectra of the 
system \cite{khan:085201}. We studied this model 
system before using several different approaches\cite{ramon:181101,ramon:21001,tamura:107402,pereverzev:104906}.
As before, we consider the initial excitonic state as being prepared in the
$|XT\rangle$ state corresponding to the photoexcitation of the F8BT
polymer.  The full set of parameters for the model may be obtained as the supplemental material of Ref.~\cite{pereverzev:104906}.

\subsection{Contributions from Coupling Modes}

\begin{figure}[t]
\subfigure[]{\includegraphics[width=0.48\columnwidth]{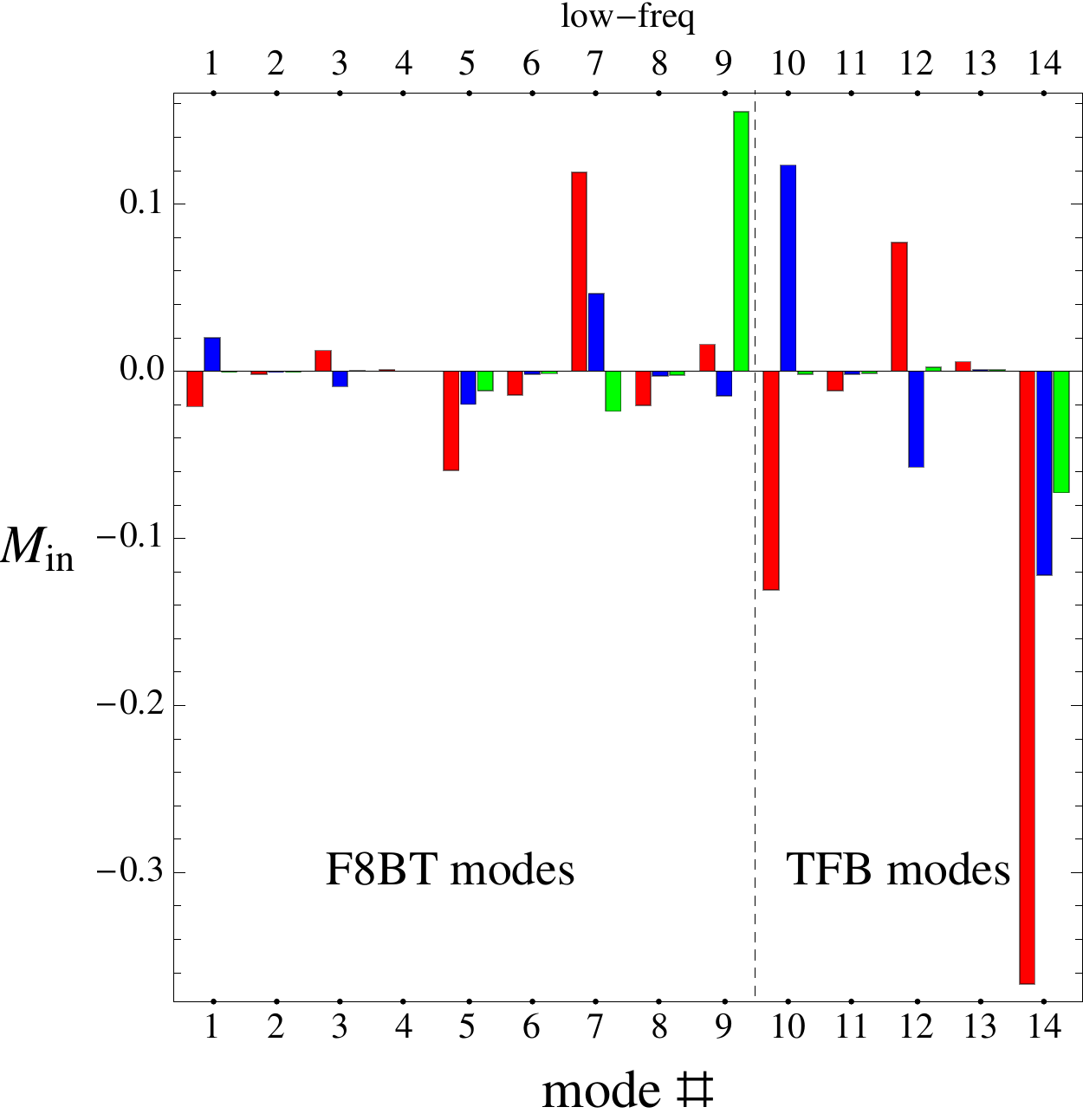}}
\subfigure[]{\includegraphics[width=0.48\columnwidth]{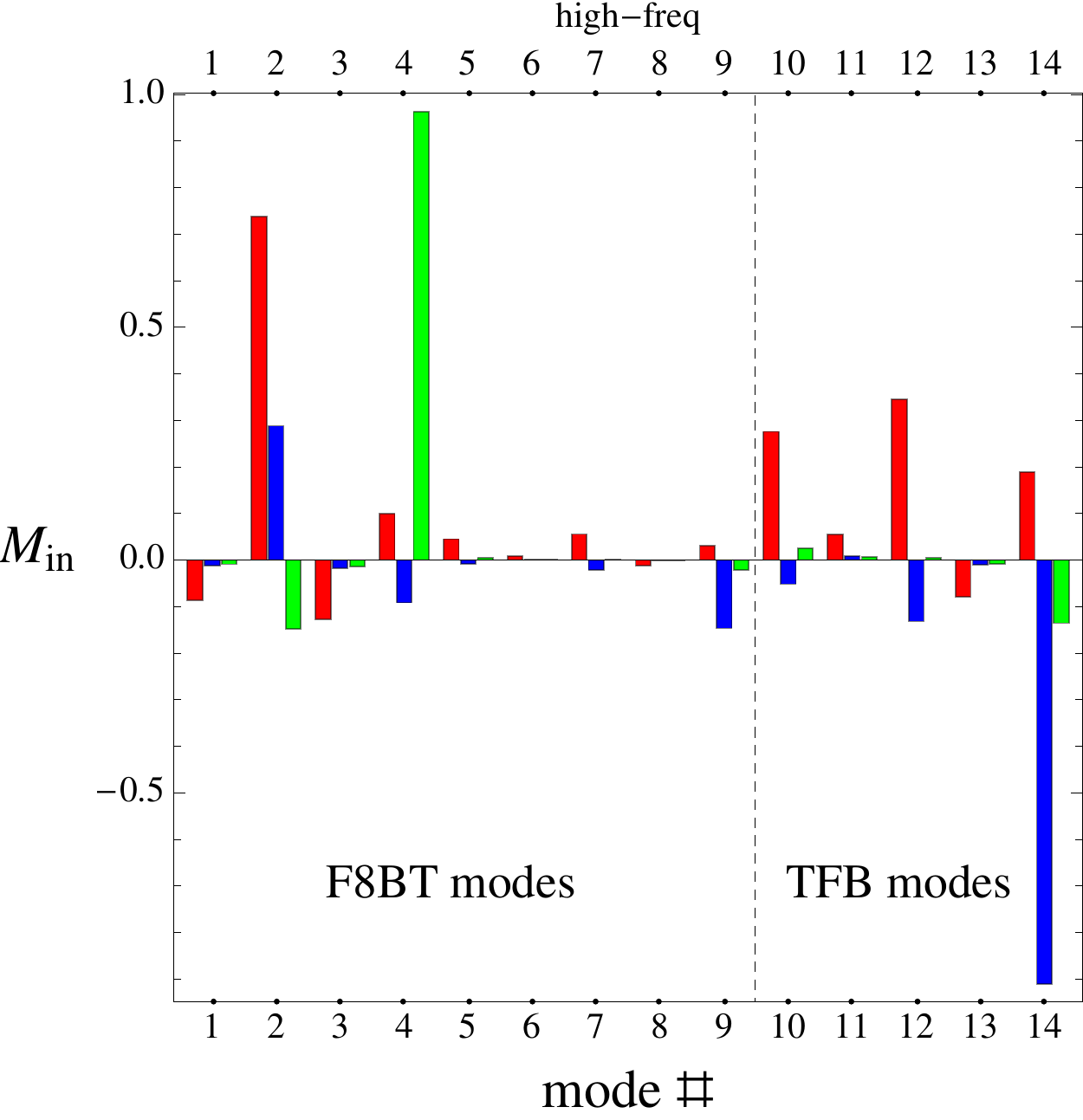}}
\caption{Projection of the three coupling modes onto the primitive modes for the model heterojunction system.  
(a)  Projection onto low frequency torsional modes and  (b)  projection on to the higher frequency C=C modes included in the model.  
Red, Blue, and Green indicate the contributions from coupling modes 1,2, and 3 respectively. }
\label{modes1}
\end{figure}

First, we examine the effect of mode-projection on the model Hamiltonian describing the 
heterojunction.  Following the mode-projection, the 28 primitive modes reduce to three 
coupling modes with frequencies $\omega_s = 1587.61 $cm$^{-1}$, 1625.07  cm$^{-1}$,
and 1637.22 cm$^{-1}$ respectively.   The projection of these modes onto the primitive basis is 
shown in Fig.~\ref{modes1}.   These projected modes involve contributions from both the high and 
low frequency phonon bands, although the dominant contributions come from the higher frequency 
phonon band.  Furthermore, since the primitive modes are localized to the individual molecular units, 
we can comment that the first coupling mode $\omega_1 =  1587.61 $ cm$^{-1}$ (shown in red in 
Fig.~\ref{modes1})  involves contributions from both molecules, while the other two modes (shown in 
blue and green, in Fig.~\ref{modes1} respectively) are dominated by local contributions coming from 
either the F8BT chain or the TFB chain.  

\begin{figure}[t]
\begin{center}
\subfigure[]{\includegraphics[width=0.40\columnwidth]{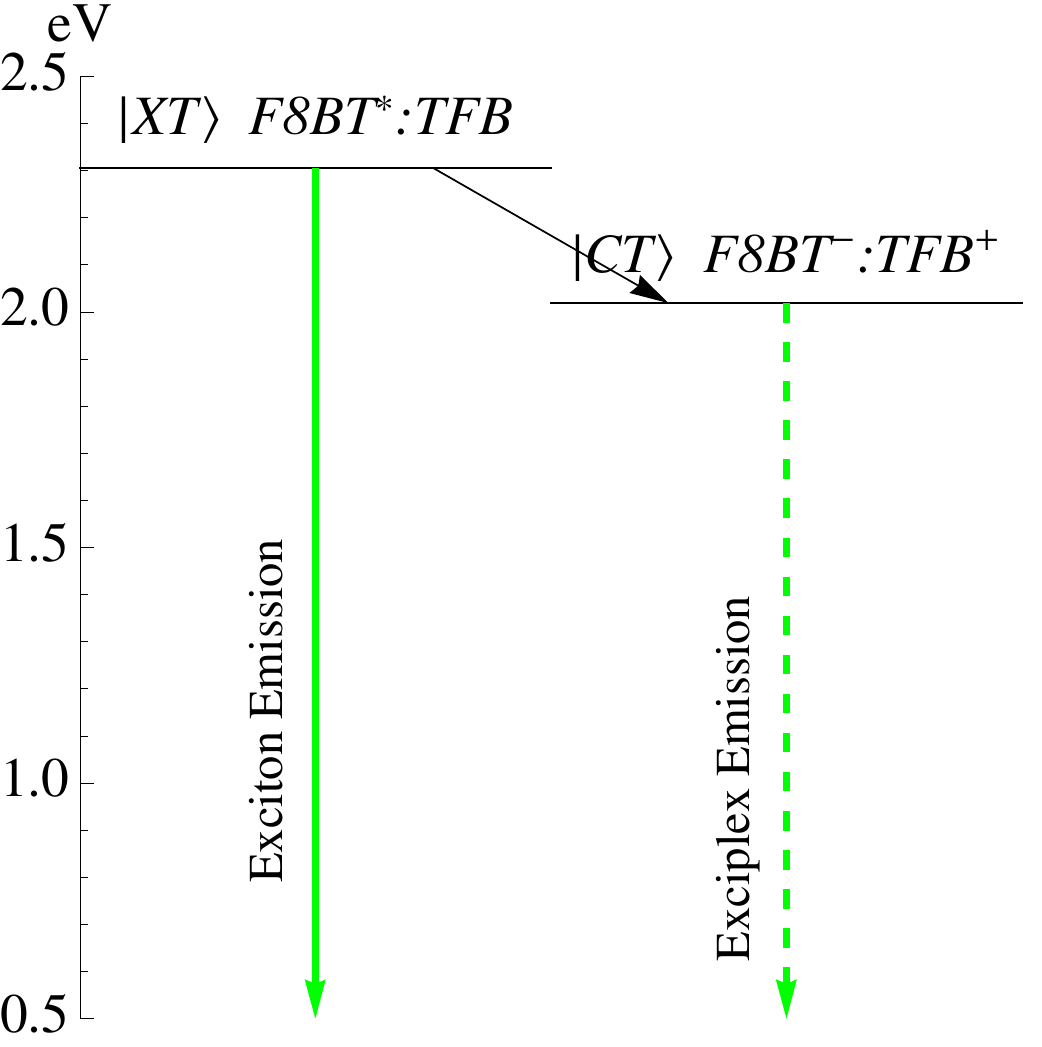}}
\subfigure[]{\includegraphics[width=0.45\columnwidth]{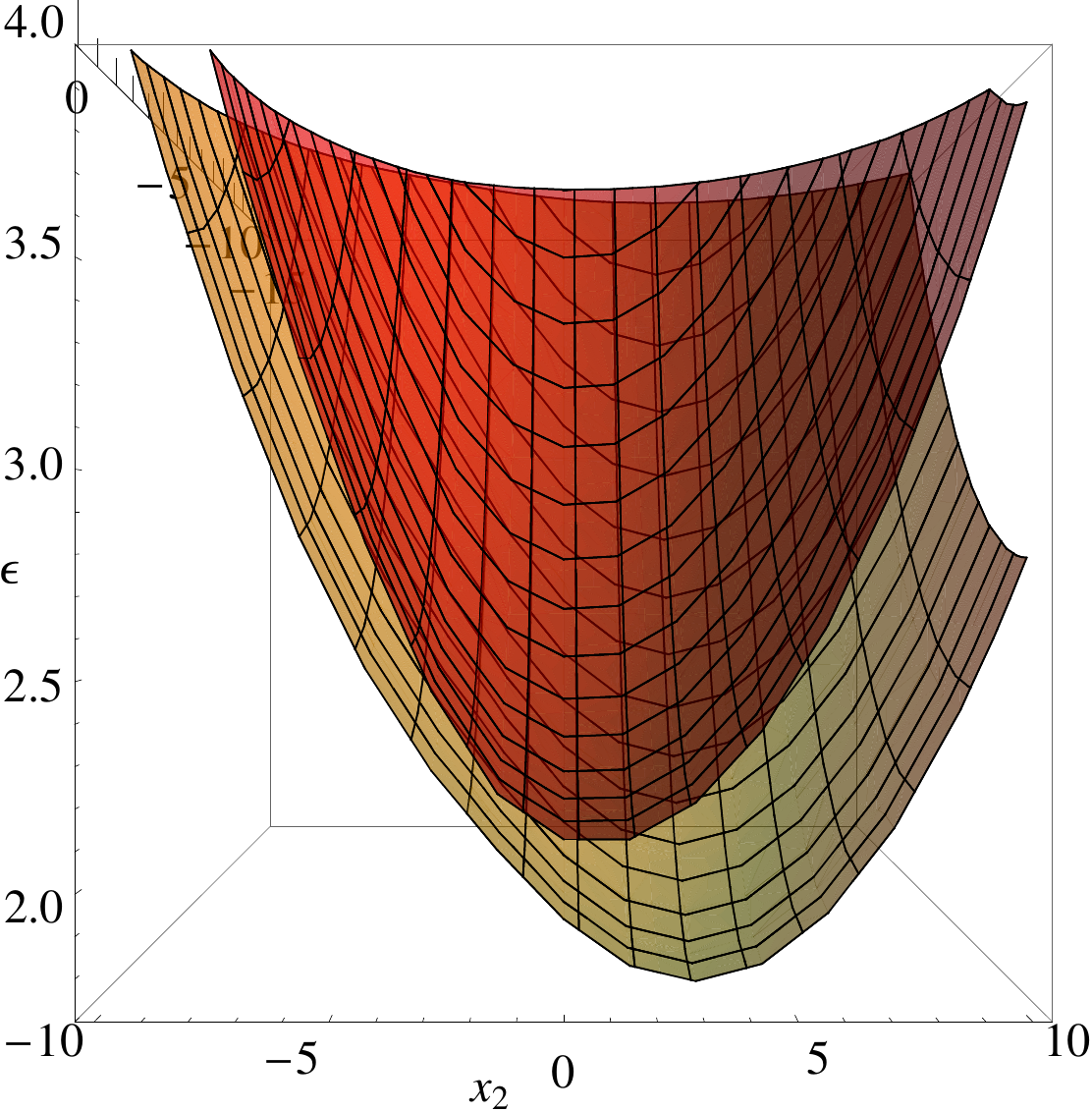}}
\end{center}
\caption{
a.) Energy level diagram showing relevant electronic states and corresponding vertical excitation energies. 
b.) A slices through the 3-dimensional Born-Oppenheimer potential energy surface for the model heterojunction dimer.  Here, 
$x_3$ is fixed at the global minima (corresponding to the CT diabatic curve), 
}\label{pes}
\end{figure}

In Fig.~\ref{pes} we show some details of the potential energy surface generated by our model Hamiltonian.  
Fig.~\ref{pes}a shows the position of the two relevant electronic states corresponding to the emissive excitonic state
and the weakly emissive exciplex state.   Within the model, there does exist an intermediate charge-transfer 
state that is energetically close to the XT state.   We recently explored the and energetics~\cite{ramon:181101}
and dynamical  \cite{tamura:107402} implications of such dark exciplex states using quantum chemical and numerically
exact quantum propagation.   In the initial transfer, the intermediate state (for this system at least) plays a minimal role 
with most of the population being transferred directly from the XT to CT. However, it does play a role in the back transfer 
process.     In Fig.~\ref{pes}b we give a slice along along one of the projected mode coordinates ($x_2$) keeping $x_3$ fixed
at the energy minimum of the CT state.  The slice where $x_1$ is at the CT energy minimum indicates that the system is nearly in the 
``barrierless'' regime for electron transfer as suggested by the experimental observations.\cite{morteani:1708}


\subsection{Population Relaxation}

We now examine the dynamical ramifications of the mode projection scheme for the system following excitation to the 
XT state.  
In Ref.~\cite{pereverzev:104906} we showed that to second-order in the electron/phonon couplings,
 non-Markovian contributions to the population decay can be included within a time-convolutionless form
 of the master equation (TCLME)
\begin{eqnarray}
\frac{d P_\alpha(t)}{d t}=\sum_\beta W_{\alpha\beta}(t)P_\beta(t)-\sum_\beta W_{\beta\alpha}(t)P_{\alpha}(t) \label{Pauli}.
\end{eqnarray}
in which the time dependent rates $W_{\alpha\beta}(t)$ are  given by
\begin{eqnarray}
W_{\beta\alpha}(t)=2 Re\int_0^td\tau\langle M_{\alpha\beta}M_{\beta\alpha}(\tau)\rangle
e^{-i(\tilde\epsilon_\alpha-\tilde\epsilon_\beta)\tau}, \label{rates}
\end{eqnarray}
where 
\begin{eqnarray}
\langle M_{\alpha\beta}M_{\beta\alpha}(\tau)\rangle=
{\rm Tr}\left(M_{\alpha\beta}M_{\beta\alpha}(\tau)\rho_{eq}\right). \label{corrfun}
\end{eqnarray} 
is the autocorrelation of the off-diagonal coupling operator in the Heisenberg representation for the canonical ensemble $\rho_{eq}$.
 The explicit form of  $C_{\alpha\beta}(\tau) = \langle M_{\alpha\beta}M_{\beta\alpha}(\tau)\rangle$ is complicated and lengthy and the reader is 
referred to the original work for its details.   
Provided that  the correlation function $C_{\alpha\beta}(t) \rightarrow 0$ at long times, the golden-rule
rate is given by 
\begin{eqnarray}
W_{\beta\alpha}^{g.r.} = \lim_{t\rightarrow \infty} W_{\beta\alpha}(t).
\end{eqnarray}
At long times when the rates become constant the TCLME preserves both positivity and detailed balance.  However, 
at intermediate times one can obtain negative rates depending upon the spectral density of the coupling\cite{pereverzev:104906}.

%

In Fig. \ref{Corr} and Fig.~\ref{fig1} we compare the results of various approximate decoupling scheme with results obtained without using the 
decoupling.   As a first approximation, we disregard all the bath modes and their couplings to the system modes. Thus, only the
terms describing  electronic degrees of freedom and the three $c$-modes are kept. 
Clearly, this can only be a good approximation if in the transformed Hamiltonian  (Eq. \ref{Ham-prime}) 
the coupling between the system modes and bath modes is weak.   For the case at hand, this proves to  be  a reasonable approximation for a short time dynamics but a poor approximation for longer times as the population never relaxes from the 
initial state.  This  indicates that the coupling between the $c$-modes and residual $r$ modes is sufficiently strong and cannot be ignored. 

\begin{figure}
 \includegraphics[width=\columnwidth]{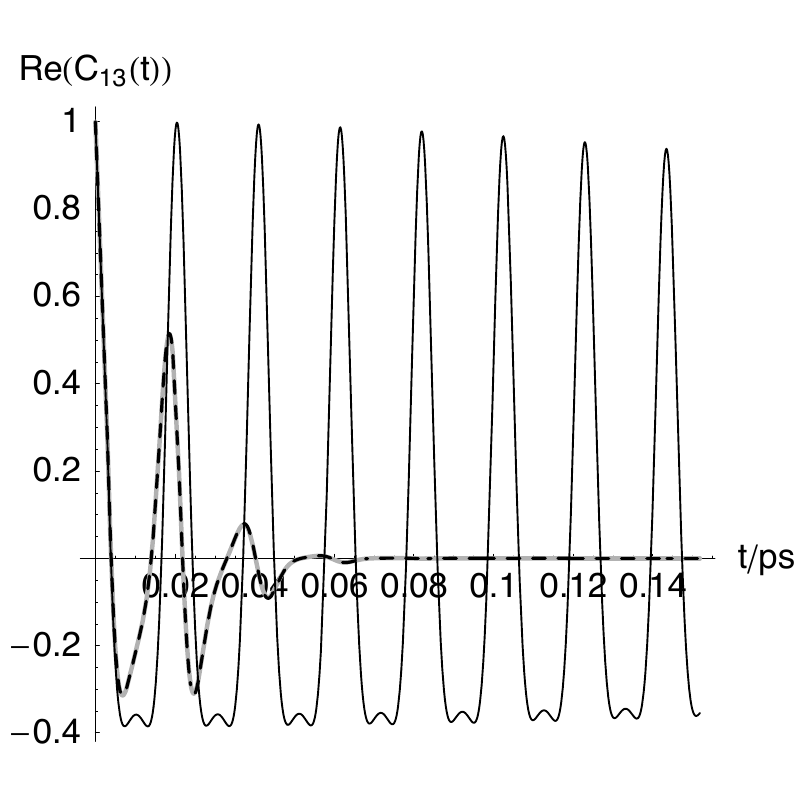}  
 \caption{\label{Corr} The real part of the correlation function $C_{13}(t)$ 
  for  the fully-coupled 28 modes  model (solid gray)
 and  reduced  3 (solid black) and 6 (dashes) mode decoupled models.  Here we use 
 a temperature of $T = 298 $K. 
}\label{Corr}
 \end{figure}

 The picture is completely different if we perform independent unitary transformations  within each of the two phonon bands of our model.
 In this case the electronic degrees of freedom are coupled to a total of six $c$-modes, three for each phonon band. 
 Within the TCLME formalism, the computed population decay is virtually indistinguishable from the fully coupled (28-mode) case even out to long times. 

\begin{figure}
 \includegraphics[width=\columnwidth]{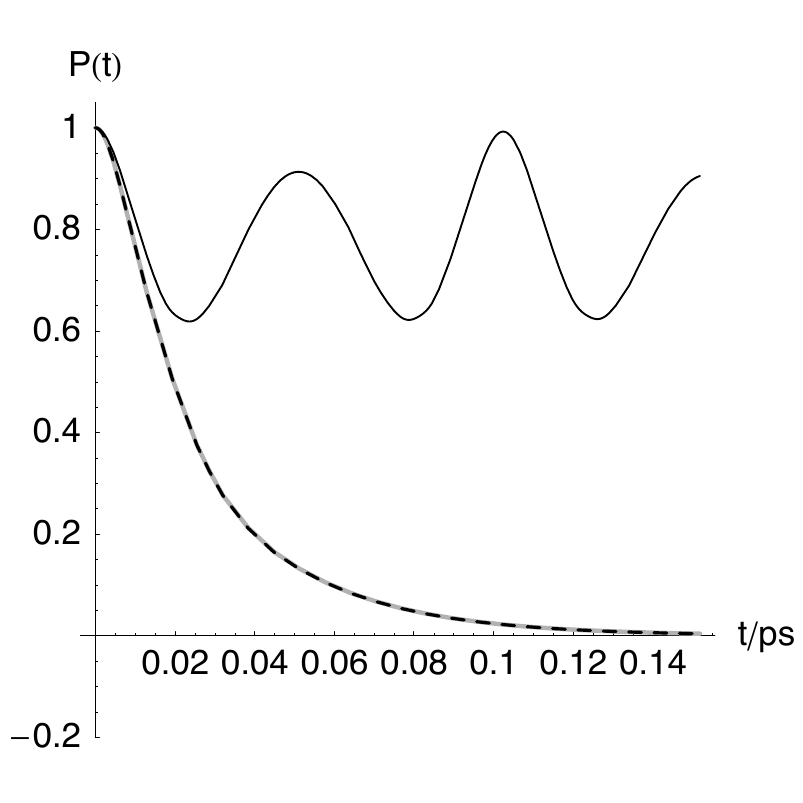}  
 \caption{\label{Orthogonal-2} The TCLME population of the excited state as a
 function of time for  the fully-coupled 28 modes  model (solid gray)
 and  reduced  3 (solid black) and 6 (dashes)  mode decoupled models.  $T = 298 $K. 
}\label{fig1}
 \end{figure}
 
In Ref. \cite{tamura:021103,Tamura:2006} the same model was investigated by using different truncations of the Mori chain with the time evolution 
of the multi-dimensional electron/phonon wavefunction was propagated 
using the numerically exact multi-configuration time-dependent Hartree  (MCTDH) method\cite{mctdh:package,beck:1}.
The results obtained there are qualitatively similar to ours in that too severe of a truncation produces no relaxation 
and that both high and low frequency contributions are needed to affect the fission of the exciton.

 \section{Discussion}
 
Here we have presented one very straightforward and essentially universal procedure for reducing the 
complexity of a quantum dynamical simulation of a multi-dimensional multi-state system by reducing the number of normal modes directly interacting with the electronic system. Splitting all the normal modes into two groups allows one to apply different levels of theory to each of these groups. In particular, one can expect that as far as the electronic dynamics goes that the modes that are directly coupled to the electronic degrees of freedom are more important and showed be treated by more exact methods while the residual modes can be treated semi-classically as a bath. Such approximations should certainly be justified if the system modes are weakly coupled to the bath modes. We showed that rather then using the minimum number of system modes it is possible to adjust the number of the system modes to reduce their coupling the bath modes. It would be desirable to develop an unambiguous procedure for choosing the optimum number of modes that are still few in number and at the same time weakly coupled to the residual modes. So far the only unambiguous criteria that we have is that doing transformations of the type of Eq. (\ref{transx}) within the groups of modes with near-degenerate frequencies generally leads to the weak system-bath mode coupling. In particular, in case of complete degeneracy, the system and bath modes are completely decoupled.

\appendix
\section{ }

Consider the linear transformations
\bea
X_i&=&\sum_jM_{ij}x_j,  \\ 
P_i&=&\sum_j (\widetilde{M}^{-1})_{ij}p_j, \label{equa}
\eea
where $M_{ij}$ are the elements of an arbitrary real invertible matrix ${\bf M}$.
In this appendix we will show that these transformations 
are equivalent to the following  unitary transformations of operators $x_i$ and $p_i$
\be
X_{i}=U^{-1}x_iU, \qquad P_{i}=U^{-1}p_iU, 
\ee
and, therefore, transformations in Eqs. (\ref{equa}) are canonical even though they are not orthogonal. 

By virtue of the singular value decomposition theorem  
any invertible real matrix $\bf M$ can be written  as
\be
{\bf M}={\bf O}_1{\bf D}\widetilde{{\bf O}}_2, \label{singular}
\ee
where ${\bf O}_1$ and ${\bf O}_2$ are orthogonal matrices and  ${\bf D}$ is a positive definite diagonal matrix.
It is sufficient to show that that there is a unitary transformation corresponding to each of these 
matrices. Then the unitary transformation corresponding to ${\bf M}$ will be a product of these
transformations.
Consider the orthogonal matrices first. They can have a determinant of either $1$ or $-1$.
If the determinant of an orthogonal matrix is positive it corresponds rotations in the multi-dimensional coordinate space.
Let us show that  in can be represented by the following unitary transformation of $x_i$ and  $p_i$
\be
X_i=U_o^{-1}x_iU_o, \qquad P_i=U_o^{-1}p_iU_o,\label{unix} 
\ee
where
\be
U_o=\exp\bigg(-\frac{i}{2}\sum_{ij}A_{ij}(p_ix_j+x_jp_i)\bigg)
\ee
and  ${\bf A}$ is a real anti-symmetric  matrix. The explicit effect of the unitary transformation 
\ref{unix} can be calculated by introducing an auxiliary operator 
\be
U_o(\xi)=\exp\bigg(-\frac{i}{2}\sum_{ij}A_{ij}(p_ix_j+x_jp_i)\xi\bigg).
\ee
Consider the auxiliary coordinate and momentum operators
\be
X_i(\xi)=U_o^{-1}(\xi)x_iU_o(\xi),\qquad P_i(\xi)=U_o^{-1}(\xi)p_iU_o(\xi).
\ee
 Differentiating $X_i(\xi)$ and $P_i(\xi)$ with respect to $\xi$ and using the commutations relations between coordinates 
 and momenta we obtain
 \be
 \frac{dX_i(\xi)}{d\xi}=\sum_jA_{ij}X_j(\xi),\qquad \frac{dP_i(\xi)}{d\xi}=\sum_jA_{ij}P_j(\xi).
 \ee 
 Solving the last two equations subject to the initial conditions $X_i(0)=x_i$ and $P_i(0)=p_i$ and bearing in mind that $X_i=X_i(1)$,
 $P_i=P_i(1)$ 
 we have
 \be
 X_i=\sum_j\left(e^A\right)_{ij}x_{j},\qquad P_i=\sum_j\left(e^A\right)_{ij}p_{j}. \label{export}
 \ee
 Since any orthogonal matrix with determinant $1$ can be written as an exponential of an anti-symmetric
 matrix  we can reproduce any orthogonal matrix in Eq. \ref{export} by a suitable choice
 of  $\bf A$.
 Thus,  any orthogonal transformation with determinant $1$  can be represented by a unitary transformation of the form
of Eq. (\ref{unix}). 
Let us now show that there exists a suitable unitary transformation of $x_i$ and $p_i$ if the determinant of 
either of the orthogonal matrices in Eq. \ref{singular} is negative. Any orthogonal matrix with
the determinant of $-1$ can be written as a product of an arbitrary orthogonal matrix with
the determinant of $-1$ and an orthogonal matrix with the determinant of $1$
\be
O_{-1}=O_{-1}\widetilde{O'}_{-1}{O'}_{-1}=O''_1{O'}_{-1}.
\ee
The case of the positive determinant orthogonal matrix was considered above. Therefore, we can choose a
simple orthogonal matrix with determinant $-1$ for ${O'}_{-1}$ and show that there is a corresponding 
unitary transformation. A simple choice is a diagonal matrix one of whose diagonal elements is equal to 
$-1$ and the rest are equal to $1$.  Its action corresponds to the space inversion for one of the oscillators
and has the following unitary realization
\be
X_i=U_q^{-1}x_iU_q, \qquad P_i=U_q^{-1}p_iU_q
\ee
with 
\be
U_q=\exp\left(i\pi\left(\frac{p^2_i}{2}+\frac{x^2_i}{2}\right)\right).
\ee
Finally matrix ${\bf D}$ being a positive definite diagonal matrix corresponds to 
the rescaling of the coordinate and momentum operators of the following form
\be
X_i=e^{B_i}x_i,\qquad P_i=e^{-B_i}p_i
\ee
and also 
has a unitary realization as 
\be
X_i=U_s^{-1}x_iU_s, \qquad P_i=U_s^{-1}p_iU_s,
\ee
where
\be
U_s=\exp\left(-i\sum_i\frac{B_i}{2}\left(x_ip_i+p_ix_i\right)\right)
\ee
Thus , we showed that transformations in Eqs. (\ref{equa}) can be viewed as unitary transformations of operators $x_i$ and $p_i$.

\section{ }

In this appendix we will show that it is possible to transform Eq. (\ref{Ham-prime}) to the form in which there are still $n_s$ system modes but each of the operators $|a\ket\bra b|+|b\ket\bra a|$ (for $a\neq b$) and $|a\ket\bra a|$ interacts with its own single mode. We consider the case of a two-level system, generalizations to higher-level systems are trivial.
Consider the case of two-level electronic system and, therefore,
three system modes ($n_s=3$). Let us show that it can be transformed into
\bea
H&=&
\sum_{a=1}^{2}\epsilon_a |a\ket\bra a|+
Y_1 |1\ket\bra 1|+Y_2 |2\ket\bra 2|
+Y_3\Big(|1\ket\bra 2|+|2\ket\bra 1|\Big)\non \\ 
& &+  
\sum_{i=1}^{3}\sum_{j=4}^N G_{ij}Y_iX_j
+\sum_{i,j=1}^{3}\frac{U_{ij}{\Pi}_i{\Pi}_j}{2}
+\sum_{i,j=1}^{3}\frac{V_{ij}{Y}_i{Y}_j}{2}  \nonumber \\
&+&\sum_{i=4}^{N}\frac{{P}_i^2}{2} 
+ \sum_{i=4}^{N}\frac{{\omega'}_i^2{X}_i^2}{2}, \label{HamY}
\eea
where we have written the electron-phonon interaction part explicitly. Thus, only one mode is 
directly responsible for the electronic transitions.
Eq. \ref{HamY} can be obtained from Eq.  \ref{Ham-prime} by the following  non-orthogonal transformation of coordinate and momentum operators in the space of system modes
\be
Y_i=\sum_{i=j=1}^3L_{ij}X_j, \qquad \Pi_i=\sum_{i=j=1}^3(\widetilde{L}^{-1})_{ij}P_j.
\ee
This is a special case of 
the coordinate transformations we used earlier. 
Here $L_{ij}$ are elements of the following $3\times3$ matrix ${\bf L}$
\be
{\bf L}=\pmatrix{g_{111}&g_{112}&g_{113}\cr g_{221}&g_{222}&g_{223}\cr g_{121}&g_{122}&g_{123}\cr}
\ee
and $(\widetilde{M}^{-1})_{ij}$ denote matrix elements of the inverse of the transpose 
of ${\bf L}$. Matrix elements $U_{ij}$ and $V_{ij}$ in Eq. \ref{HamY} are given by 
\be
U_{ij}=\sum_{k=1}^3\widetilde{L}_{ki}\widetilde{L}_{kj},\qquad V_{ij}=\sum_{k=1}^3{\omega'_k}^2(L^{-1})_{ki}(L^{-1})_{kj}, 
\ee
and 
\be
G_{ij}=\sum_{k=1}^3\gamma'_{kj}(L^{-1})_{ki}.
\ee
We would like to stress again here that even though matrix ${\bf L}$ is non-orthogonal operators $Y_i$ and $\Pi_i$ are unitarily equivalent to, respectively, $X_i$ and $P_i$, and
therefore, Hamiltonian (\ref{HamY}) is unitarily equivalent to Hamiltonian (\ref{Ham-prime}).

\begin{acknowledgments}
This work was funded in part through grants from the National Science
Foundation (CHE-0712981) and the Robert A. Welch foundation (E-1337).
Support from the Texas Learning and Computational Center (TLC$^2$) is also greatly acknowledged.
The authors also wish to thank Dr. Irene Burghardt (ENS/Paris) for numerous fruitful discussions 
regarding this and related works. \end{acknowledgments}



\end{document}